\documentstyle[aps]{revtex}
\tighten

\begin{document}

%======================================%
%<<<<<<<<<<<< TITLE PAGE >>>>>>>>>>>>>>%
%======================================%

\preprint{YITP-00-25, KUCP0152, gr-qc/0005xxx}

\title{Excitation of Kaluza-Klein modes of $U(1)$ field by parametric
resonance} 
\author{Yoshiyuki Morisawa}
\address{
Yukawa Institute for Theoretical Physics, Kyoto University\\
Kyoto 606-8502, Japan}
\author{Kunihito Uzawa}
\address{
Graduate School of Human and Environment Studies, Kyoto University\\
Kyoto 606-8501, Japan}
\author{Shinji Mukohyama}
\address{
Department of Physics and Astronomy, University of Victoria\\
Victoria, BC, V8W 3P6, Canada}
\date{\today}

\maketitle

%======================================%
%<<<<<<<<<<<<< ABSTRACT >>>>>>>>>>>>>>>% 
%======================================%

\begin{abstract} 

In this paper, based on the theory of parametric resonance, we
propose a cosmological criterion on ways of compactification: 
we rule out such a model of compactification that there exists a
Kaluza-Klein mode satisfying $2m_{KK}=\omega_b$, where $m_{KK}$ is
mass of the Kaluza-Klein mode and $\omega_b$ is the frequency of
oscillation of the radius of the compact manifold on which extra
dimensions are compactified. This is a restatement of the criterion
proposed previously 
[S. Mukohyama, Phys. Rev. {\bf D57}, 6191 (1998)]. 
As an example, we consider a model of compactification by a sphere and
investigate Kaluza-Klein modes of $U(1)$ field. In this case the
parametric resonance is so mild that the sphere model is not ruled
out. 

\end{abstract}


%======================================%
%<<<<<<<<<<< Introduction >>>>>>>>>>>>>%
%======================================%
\section{Introduction}

%%% Higher dimensional theory

Many unified theories, such as superstring theory~\cite{Superstring}
and M-theory~\cite{M-theory}, require the dimensionality of spacetime 
larger than four. In these theories, we have to reduce spacetime
dimensions to obtain a four-dimensional theory describing the
observed universe. While a new mechanism was recently proposed by
Randall and Sundrum~\cite{RS} and has been attracting much interests,
the Kaluza-Klein prescription of compactification~\cite{KK} still has
been conventionally adopted. In particular, combination of these two
mechanism seems promising. In the Kaluza-Klein prescription, extra
dimensions are compactified on a compact manifold not to be seen 
in low energy. As we can easily expect, the corresponding
four-dimensional effective theory depends strongly on the way of
compactification. This may be considered as an advantage: the
four-dimensional theory including abundant contents may emerge from a
simple higher-dimensional theory as a result of dynamics of the
theory. However, at the same time, it may be considered as a
disadvantage: we cannot give definite predictions in four-dimension
without specifying the way of compactification.

%%% Kolb & Slansky

In this respect, Kolb and Slansky~\cite{Kolb&Slansky} did an
interesting work. They considered a five-dimensional theory including
gravity and a massless scalar field. After compactifying the extra
one-dimension by a circle, they investigated cosmological evolution of 
energy density $\rho_{KK}$ of Kaluza-Klein modes of the scalar field,
provided that entropy production is negligible. It was concluded that,
if $\rho_{KK}$ is comparable to radiation energy density $\rho_{rad}$
at an early epoch of the universe, the former density exceeds the
critical density of the universe soon. Therefore, quanta of
Kaluza-Klein modes must not be excited catastrophically in the early
universe.

Their conclusion is essentially because of the momentum conservation
along the circle. Because of it, a quantum of a Kaluza-Klein mode
cannot decay into zero modes without meeting accidentally with a
quantum of another Kaluza-Klein mode with the exactly opposite
momentum along the circle. Hence, $\rho_{KK}$ evolves simply like
$\propto a^{-3}$, and $\rho_{KK}/\rho_{rad}\propto a$, where $a$ is
the scale factor of the universe.

%%% Generalization of Kolb & Slansky

Since the momentum conservation along the compact manifold is expected
to hold in a large class of models of compactification, it seems that
the above result can be generalized. Namely, {\it quanta of
Kaluza-Klein modes must not be excited so catastrophically that energy 
density of them become comparable to radiation energy density}.

%%% Purpose of this paper

Purpose of this paper is to analyze whether quanta of Kaluza-Klein
modes are excited by parametric resonance caused by small oscillation
of the radius of compactification. Such a analysis may give a
cosmological criterion on ways of compactification. In
Sec.~\ref{sec:criterion} we review and restate clearly the
cosmological criterion proposed in ref.~\cite{Mukohyama}. In
Sec.~\ref{sec:U(1)}, based on the criterion, we consider a model of
compactification by a sphere and investigate Kaluza-Klein modes of a
$U(1)$ field. We conclude that excitation of Kaluza-Klein modes is so
mild that the sphere model is not ruled out. Sec.~\ref{sec:summary} is 
devoted to a summary of this paper. In Appendix~\ref{app:proca} we
derive hamiltonian for a massive vector field (or a Proca field) for
the use in Sec.~\ref{sec:U(1)}.

%======================================%
%<<<<<< Cosmological criterion>>>>>>>>>%
%======================================%
\section{Cosmological criterion on compactification}
	\label{sec:criterion}

%%% Situation

Let us suppose the following situation: 
(a) we consider a $D$-dimensional theory which includes gravity and
other fields;
(b) we adopt the conventional Kaluza-Klein compactification as a way
of dimensional reduction: we compactify the extra $(D-4)$-dimensions
on a compact manifold;
(c) we consider some mechanism (e.g. the Casimir effect) to stabilize
the compactification;
(d) we also consider perturbations of $D$-dimensional fields around
the background specified by the Kaluza-Klein prescription. 

%%% Oscillation of mass

The perturbations in (d) include Kaluza-Klein modes which can be
considered as massive fields in four-dimension. Squared mass of the
Kaluza-Klein modes are essentially eigenvalues of the Laplacian on
the compact manifold. Hence, it is proportional to $b^{-2}$, where $b$
is the radius of the compact manifold. On the other hand, (c) implies
that $b$ has a potential with at least one local minimum, say at
$b=b_0$, when it is considered as a four-dimensional field. Evidently,
$b_0$ can be considered as the present value of $b$. Thus, it is
natural to assume that $b$ oscillates around $b_0$ in the early
universe. The frequency of the oscillation is expected to be of the
order of $b_0^{-1}$. Therefore, it is expected that in the early
universe the mass of the Kaluza-Klein modes, which is of the order of 
$b_0^{-1}$, also oscillates with the frequency of the order of
$b_0^{-1}$. Although there are freedom of conformal transformation and
field redefinitions which will be explained in the next section for a
particular model, the above expectation still seems to hold for a
large class of compactification. Namely, we can expect that {\it in
the early universe the mass of the Kaluza-Klein mode oscillates with
the frequency of the same order as the mass itself}. In such a
situation, parametric resonance might occur to enhance creation of
quanta of the Kaluza-Klein mode.

%%% Hamiltonian for KK mode

In order to analyze the parametric resonance phenomenon in the
expanding universe, we need hamiltonian of each Kaluza-Klein
mode. Considering the FRW spacetime as the four-dimensional spacetime
and performing Fourier transformation (or harmonic expansion)
w.r.t. the three-dimensional space, hamiltonian for each Fourier
component of a Kaluza-Klein mode is expected to have the following
form. 
%============< EQUATION >==============%
%
\begin{equation}
 H = \frac{1}{2}[P^2+\Omega^2(t)Q^2], 
\end{equation}
%======================================%
where $Q$ is a Fourier component of a Kaluza-Klein mode, $P$ is its
conjugate momentum, and 
%============< EQUATION >==============%
%
\begin{equation}
 \Omega^2(t) = m_{KK}^2[1-\epsilon\cos(\omega_b t)] + a^{-2}k^2
	+ O(\epsilon^2).
\end{equation}
%======================================%
Here, the constants $m_{KK}$, $\omega_b$ and $k^2$ are the mass of the 
Kaluza-Klein mode at $b=b_0$, the frequency of the oscillation of $b$
and the squared comoving momentum along the three-dimensional space,
respectively. The scale factor of the FRW universe was written as $a$, 
and $\epsilon$ is a dimensionless quantity proportional to the
amplitude of the oscillation of $b$. Throughout this paper we assume
that $\epsilon\ll 1$, which will be derived from another more essential
assumption. 

%%% Parametric resonance

From this form of the hamiltonian, the following is easily
expected:
if $A+B\approx 1$ then the parametric resonance occurs to enhance the 
process $b\to KK+\overline{KK}$; 
if $A+B\approx 4$ then the parametric resonance enhances the process
$b\times 2\to KK+\overline{KK}$; $\cdots\cdots$; 
if $A+B\approx n^2$ then the process $b\times n\to KK+\overline{KK}$
is enhanced, where $A$ and $B$ are defined by
%============< EQUATION >==============%
%
\begin{eqnarray}
 A & = & 4\cdot\frac{a^{-2}k^2}{\omega_b^2},	\nonumber\\
 B & = & 4\cdot\left(\frac{m_{KK}}{\omega_b}\right)^2,
\end{eqnarray}
%======================================%
and $b$, $KK$ and $\overline{KK}$ represent a quanta of the
oscillation of the field $b$, a quanta of a Kaluza-Klein mode and a
quanta of another Kaluza-Klein mode with the exactly opposite momentum
along the compact manifold, respectively. This observation follows
from the fact that $A+B$ is essentially equal to
$(2E_{KK}/\omega_b)^2$, where $E_{KK}$ is energy of the Kaluza-Klein
mode. Note that $A$ depends on time while $B$ does not. In particular,
$\dot{A}/A=-2H$, where $H\equiv\dot{a}/a$ is the Hubble
parameter.

%%% Creation rate and number of created quanta

The creation rate of quanta of the Kaluza-Klein mode is given by 
%============< EQUATION >==============%
%
\begin{equation}
 \Gamma \equiv \frac{d}{dt}\sinh^{-1}\sqrt{N_{KK}}
	\sim \epsilon^n\omega_b \sim \frac{\epsilon^n}{b_0}
	\quad\mbox{for}\ 
	\left|(A+B)-[n^2+O(\epsilon^2)]\right|<O(\epsilon^n)
	\ (n=1,2),
	\label{eqn:creation-rate}
\end{equation}
%======================================%
and otherwise $\Gamma \equiv (d/dt)\sinh^{-1}\sqrt{N_{KK}}\sim
0$, where $N_{KK}$ is the number of created quanta~\cite{Mukohyama}. 
By using this formula, we can estimate $N_{KK}$. For a mode satisfying 
$(n-1)^2<B<n^2$ and $B-n^2=O(1)$, the duration $\Delta t$ of the
parametric resonance is estimated to be $\sim H^{-1}\epsilon^n$. Thus, 
integration of (\ref{eqn:creation-rate}) gives 
%============< EQUATION >==============%
%
\begin{equation}
 \sinh^{-1}\sqrt{N_{KK}} \sim \Gamma\cdot\Delta t
	\sim \frac{\epsilon^{2n}}{Hb_0}
	\sim \epsilon^{2n-1}\sqrt{\frac{\rho_b}{\rho_0}},
	\label{eqn:NKK1}
\end{equation}
%======================================%
where $\rho_{b}\sim\epsilon^2b_0^{-2}/\kappa^2$ and 
$\rho_0\equiv 3H^2/\kappa^2$ are energy density of the oscillation of
$b$ and the critical density of the universe, respectively. Here,
$\kappa^2$ is the four-dimensional gravitational coupling constant. To 
obtain this estimate, we used the adiabatic approximation, assuming
that $Hb_0\ll 1$ and $\dot{H}b_0^2\ll 1$. Note that the previous
assumption $\epsilon\ll 1$ can be derived from the assumption 
$Hb_0\ll 1$, since $\rho_b/\rho_0\sim (\epsilon/Hb_0)^2$. On the other 
hand, for $B=n^2$, the duration $\Delta t$ of the parametric resonance 
is $\sim H^{-1}$, and thus the integration of
(\ref{eqn:creation-rate}) gives 
%============< EQUATION >==============%
%
\begin{equation}
 \sinh^{-1}\sqrt{N_{KK}} \sim \Gamma\cdot\Delta t
	\sim \frac{\epsilon^{n}}{Hb_0}
	\sim \epsilon^{n-1}\sqrt{\frac{\rho_b}{\rho_0}}. 
	\label{eqn:NKK2}
\end{equation}
%======================================%
We used the adiabatic approximation to obtain this result, too. 
Since $\rho_b\leq\rho_0$, the r.h.s.'s of (\ref{eqn:NKK1}) and 
(\ref{eqn:NKK2}) are bounded from above by $O(\epsilon^{2n-1})$ and
$O(\epsilon^{n-1})$, respectively. Therefore, $N_{KK}$ can be of order
unity if and only if $B=1$.

%%% Created energy density 

Hence, let us estimate created energy density for the case
$B=1$. First, by definition of $A$ the typical value of $k^2$ in the
resonance band is $\sim \Delta A\cdot a^2\omega_b^2$, where 
$\Delta A$ is the band width $\sim\epsilon$. Next, the energy density
$\rho_{KK}$ of created quanta of the Kaluza-Klein mode is 
$\sim m_{KK}(k^2)^{3/2}N_{KK}$. Thus, 
%============< EQUATION >==============%
%
\begin{equation}
 \frac{\rho_{KK}}{\rho_b} \sim 
	\left(\frac{\kappa}{b_0}\right)^2\epsilon^{-1/2}N_{KK}.
	\label{eqn:rhoKK}
\end{equation}
%======================================%

%%% Backreaction

Now let us consider backreaction, which we have not yet taken into
account. First, if we take (\ref{eqn:NKK2}) and (\ref{eqn:rhoKK}) at
their face value then they imply that $\rho_{KK}\gg \rho_b$, provided
that $\rho_b\sim\rho_0$ initially and that $b_0$ is of Planck order
$\sim\kappa$. Next, the backreaction becomes important when and only
when $\rho_{KK}$ becomes comparable to $\rho_b$. Hence, this rather
overestimated result obtained without considering backreaction meant
that $\rho_{KK}\sim \rho_b$ ($\sim\rho_{rad}$) if the backreaction was
taken into account. Therefore, from the argument in the previous
section, we have to rule out models of compactification in which there
exists a Kaluza-Klein mode with $B=1$, which is equivalent to
$2m_{KK}=\omega_b$.

%%% Interaction among KK modes

In the above argument we have ignored interaction among Kaluza-Klein
modes. This treatment can be justified, provided that coupling
constants are at most of order unity in the unit of $b_0$. Actually,
in this case we can expect that interaction should be small unless
$\rho_{KK}\sim b_0^{-4}$ ($\gg\rho_{b}$). Therefore, the result
obtained above without considering interaction is not altered by
inclusion of interaction.

%%% Cosmological criterion

Finally, we propose the following cosmological criterion on models of
compactification: {\it we rule out such a model of compactification
that there exists a Kaluza-Klein mode satisfying 
%============< EQUATION >==============%
%
\begin{equation}
 2m_{KK} = \omega_b,
\end{equation}
%======================================%
where $m_{KK}$ is mass of the Kaluza-Klein mode and $\omega_b$ is the
frequency of oscillation of the radius of the compact manifold on
which the extra dimensions are compactified}.

%======================================%
%<<<<<< Analysis of $U(1)$ field >>>>>>%
%======================================%
\section{Analysis of $U(1)$ field}
	\label{sec:U(1)}

%%% Model

In this section we analyze Kaluza-Klein modes of $U(1)$ field
$\bar{A}_M$ described by the action 
%============< EQUATION >==============%
%
\begin{equation}
 I_{U(1)} = -\frac{1}{4}\int d^Dx
	\sqrt{-\bar{g}}\bar{F}_{MN}\bar{F}^{MN},
\end{equation}
%======================================%
where $\bar{F}_{MN}=\partial_M\bar{A}_N-\partial_N\bar{A}_M$. As the
compact manifold on which the extra $(D-4)$-dimensions are
compactified, we adopt the $d$-dimensional sphere ($d=D-4$):
%============< EQUATION >==============%
%
\begin{equation}
 \bar{g}_{MN}dx^Mdx^N = 
	\left(\frac{b}{b_0}\right)^{-d}g_{\mu\nu}dx^{\mu}dx^{\nu}
	+ b^2\Omega_{ij}dx^idx^j,
\end{equation}
%======================================%
where the four-dimensional metric $g_{\mu\nu}$ and the radius $b$
depend only on the four dimensional coordinates $\{x^{\mu}\}$, and
$\Omega_{ij}dx^idx^j$ is the line element of the unit $d$-sphere.

%%% Four-dimensional action

To analyze the $U(1)$ field in the background spacetime, it is
convenient to expand the field in terms of harmonics on the
$(D-4)$-dimensional sphere. 
%============< EQUATION >==============%
%
\begin{equation}
 \bar{A}_Mdx^M = b_0^{-d/2}\left[ U_{\mu}Y dx^{\mu} 
	+ b_0(\phi_T V_{(T)i}+\phi_L V_{(L)i}) dx^i \right],
\end{equation}
%======================================%
where $U_{\mu}$, $\phi_T$ and $\phi_L$ depend only on the
four-dimensional coordinates $\{x^{\mu}\}$; $Y$ and $V_{(T)i}$ are the
scalar harmonics and the transverse-traceless vector harmonics, and
$V_{(L)i}=\partial_iY$. (See Appendix of ref.~\cite{UMM} for
definitions and properties of these harmonics.) Hereafter, we omit the
summations w.r.t. eigenvalues and in eigenstates. By substituting the
expansion into the action, we obtain the corresponding decomposition
of the action $I_{U(1)}=I_T+I_U$, where 
%============< EQUATION >==============%
%
\begin{eqnarray}
 I_{T} & = & -\frac{1}{2}\int dx^4\sqrt{-g}
	\left[\left(\frac{b}{b_0}\right)^{-2}g^{\mu\nu}
	\partial_{\mu}\phi_T\partial_{\nu}\phi_T
	+ \left(\frac{b}{b_0}\right)^{-(d+4)}
	\frac{l(l+d-1)+d-2}{b_0^2}\phi_T^2\right],
	\nonumber\\
 I_U & = & -\int dx^4\sqrt{-g}\left[
	\frac{1}{4}\left(\frac{b}{b_0}\right)^d
	g^{\mu\rho}g^{\nu\sigma}F_{\mu\nu}F_{\rho\sigma}
	+ \frac{1}{2}\left(\frac{b}{b_0}\right)^{-2}
	\frac{l(l+d-1)}{b_0^2}g^{\mu\nu}U_{\mu}U_{\nu}\right], 
	\label{eqn:expanded-action}
\end{eqnarray}
%======================================%
and $F_{\mu\nu}=\partial_{\mu}U_{\nu}-\partial_{\nu}U_{\mu}$. Note
that $\phi_L$ represents gauge degrees of freedom and that $I_{U(1)}$
does not depend on it.

%%% Result

From (\ref{eqn:expanded-action}), we can see that masses $m_T$ and
$m_U$ for the four-dimensional fields $\phi_T$ and $U_{\mu}$ are given 
by 
%============< EQUATION >==============%
%
\begin{eqnarray}
 m_T^2 = \frac{l(l+d-1)+d-2}{b_0^2}, \nonumber\\
 m_U^2 = \frac{l(l+d-1)}{b_0^2}.
\end{eqnarray}
%======================================%
(See Appendix~\ref{app:proca} for a systematic treatment of the field
$U_{\mu}$ in the FRW universe.)
On the other hand, if we assume that the compactification is
stabilized by the Casimir effect~\cite{Candelas&Weinberg}, then $d$
should be larger than $1$ and the frequency $\omega_b$ of the
oscillation of $b$ is equal to $2\omega$ in
ref.~\cite{Mukohyama,UMM}. Hence, $\omega_b^2=2(d-1)/b_0^2$, and 
%============< EQUATION >==============%
%
\begin{eqnarray}
 \left(\frac{2m_T}{\omega_b}\right)^2 & = & 
	\frac{2[l(l+d-1)+d-2]}{d-1} > 1 \quad (l\ge 1), 
	\nonumber\\
 \left(\frac{2m_U}{\omega_b}\right)^2 & = & 
	\frac{2l(l+d-1)}{d-1} > 1 \quad (l\ge 1).
\end{eqnarray}
%======================================%

Finally, from the argument in the previous section, we conclude that
the parametric resonance of Kaluza-Klein modes of the $U(1)$ field is
not so catastrophic. Hence, the analysis of this section does not rule 
out the model of compactification by the $d$-dimensional sphere with
the Casimir effect.

%======================================%
%<<<<<< Summary and discussions >>>>>>>%
%======================================%
\section{Summary and discussions}
	\label{sec:summary}

%%% Summary 

In summary, we have analyzed parametric resonance of Kaluza-Klein
modes and investigated whether a model of compactification can survive
or not. 
First, based on the theory of parametric resonance, we have
restated the cosmological criterion proposed in ref.~\cite{Mukohyama}
in the following form: 
{\it we rule out such a model of compactification
that there exists a Kaluza-Klein mode satisfying 
%============< EQUATION >==============%
%
\begin{equation}
 2m_{KK} = \omega_b,
\end{equation}
%======================================%
where $m_{KK}$ is mass of the Kaluza-Klein mode and $\omega_b$ is the
frequency of oscillation of the radius of the compact manifold on
which the extra dimensions are compactified}. 
Next, as an example, we have considered a model of compactification by
a sphere and investigated Kaluza-Klein modes of $U(1)$ field. We have
concluded that parametric resonance is so mild that the sphere model
is not ruled out.

%%% Broad resonance

In order to obtain the above criterion, we have assumed that 
$Hb_0\ll 1$ and $\dot{H}b_0\ll 1$. In future works, these assumptions
should be removed, and then we should inevitably consider the broad
resonance regime $\epsilon\sim 1$~\footnote{
In this respect, ref.~\cite{Tsujikawa} may be considered as the first 
step.}. In this regime, the so called stochastic resonance
occurs~\cite{Reheating}, and the structure of resonance depends not
only on the curvature of the potential of the field $b$ at a local
minimum (or $\omega_b$) but also on the whole shape of the potential.

%%% Other KK modes

Although we have concluded that Kaluza-Klein modes of the $U(1)$ field 
are not excited catastrophically by parametric resonance and that the 
sphere model of compactification is not ruled out, still there may be
possibilities that Kaluza-Klein modes of other fields might be excited 
strongly by parametric resonance. In this respect, in
ref.~\cite{Mukohyama,UMM}, Kaluza-Klein modes of a massless scalar
field and a part of Kaluza-Klein modes of gravitational perturbations
were analyzed. It was concluded that parametric resonance of those 
Kaluza-Klein modes is also mild.

%%%%%%%%%%%%%%%%%%%%%%%%%%%%%%%%%%%%%%%%%%%%%%%%%%%%%%%%%%%%%%%%%%%%
%%%%%%%%%%%%%%%%%%%%%%%%%%%%%%%%%%%%%%%%%%%%%%%%%%%%%%%%%%%%%%%%%%%%
% Acknowledgements
%%%%%%%%%%%%%%%%%%%%%%%%%%%%%%%%%%%%%%%%%%%%%%%%%%%%%%%%%%%%%%%%%%%%
%%%%%%%%%%%%%%%%%%%%%%%%%%%%%%%%%%%%%%%%%%%%%%%%%%%%%%%%%%%%%%%%%%%%
\begin{acknowledgments}

YM would like to thank Professor T.~Nakamura for continuing
encouragement.
KU would like to thank Professor J. Soda for continuing encouragement.
SM would like to thank Professor W. Israel for continuing
encouragement. SM's work is supported by the CITA National Fellowship
and the NSERC operating research grant. 

\end{acknowledgments}

%%%%%%%%%%%%%%%%%%%%%%%%%%%%%%%%%%%%%%%%%%%%%%%%%%%%%%%%%%%%%%%%%%%%
%%%%%%%%%%%%%%%%%%%%%%%%%%%%%%%%%%%%%%%%%%%%%%%%%%%%%%%%%%%%%%%%%%%%
% Appendix
%%%%%%%%%%%%%%%%%%%%%%%%%%%%%%%%%%%%%%%%%%%%%%%%%%%%%%%%%%%%%%%%%%%%
%%%%%%%%%%%%%%%%%%%%%%%%%%%%%%%%%%%%%%%%%%%%%%%%%%%%%%%%%%%%%%%%%%%%

\appendix

%======================================%
%<<<<<<<<<<<< APPENDIX A >>>>>>>>>>>>>>%
%======================================%
\section{Proca field in FRW universe}
	\label{app:proca}

%%% Action

In this appendix we consider a massive vector field (or a Proca field)
$U_{\mu}$ in the $(n+1)$-dimensional FRW universe. The action is 
%============< EQUATION >==============%
%
\begin{equation}
 I = -\int dx^{n+1}\sqrt{-g}\left[
	\frac{f}{4}g^{\mu\rho}g^{\nu\sigma}F_{\mu\nu}F_{\rho\sigma}
	+ \frac{m^2}{2}g^{\mu\nu}U_{\mu}U_{\nu}\right], 
\end{equation}
%======================================%
where $F_{\mu\nu}=\partial_{\mu}U_{\nu}-\partial_{\nu}U_{\mu}$, and
the background is given by 
%============< EQUATION >==============%
%
\begin{eqnarray}
 g_{\mu\nu}dx^{\mu}dx^{\nu} & = & 
	-dt^2 + a^2(t)\Sigma_{pq}dx^pdx^q,
	\nonumber\\
 f & = & f(t),\nonumber\\
 m^2 & = & m^2(t).
\end{eqnarray}
%======================================%
Here, $\Sigma_{pq}dx^pdx^q$ is the line element of a $n$-dimensional
space of constant curvature, whose curvature constant is denoted by
$K$ in the following argument. By substituting these background into
the action, we obtain 
%============< EQUATION >==============%
%
\begin{eqnarray}
 I & = & \frac{1}{2}\int dtdx^n\sqrt{\Sigma}\left[ af\Sigma^{pq}
	(\dot{A}_p-\partial_pA_0)(\dot{A}_q-\partial_qA_0)
	+ a^3m^2A_0^2 \right.\nonumber\\
	& & \left. - \frac{f}{a}
	(\Sigma^{pp'}\Sigma^{qq'}-\Sigma^{pq'}\Sigma^{qp'})
	\partial_pA_q\partial_{p'}A_{q'}
	- am^2\Sigma^{pq}A_pA_q\right],
\end{eqnarray}
%======================================%
where the dot denotes a time derivative. Since the action does not
include $\dot{A}_0$, $\delta I/\delta A_0=0$ gives a
constraint. Actually, the constraint is the second class and can be
solved formally as 
%============< EQUATION >==============%
%
\begin{equation}
 U_0 = (\Delta-f^{-1}a^2m^2)^{-1}\Sigma^{pq}\partial_p\dot{U}_q,
	\label{eqn:U0}
\end{equation}
%======================================%
where $\Delta$ is the Laplacian in the $n$-dimensional space of
constant curvature. This can be used to eliminate $A_0$ from the
action.

In order to obtain explicit expression for the r.h.s. of
(\ref{eqn:U0}), we expand the field $U_{\mu}$ in terms of harmonics on 
the $n$-dimensional space of constant curvature. 
%============< EQUATION >==============%
%
\begin{equation}
 U_{\mu}dx^{\mu} = u_0ydt 
	+ (u_{\perp}v_{(T)p}+u_{\parallel}v_{(L)p})dx^p,
\end{equation}
%======================================%
where $u_0$, $u_{\perp}$ and $u_{\parallel}$ depend only on the time
variable $t$; $y$ and $v_{(T)p}$ denote the scalar harmonics and the
transverse-traceless vector harmonics, respectively, and
$v_{(L)p}=\partial_py$. Hereafter we omit the summations (or
integrations) w.r.t. eigenvalues and in eigenspaces. (For definition
of harmonics, we adopt those given in Appendix of
ref.~\cite{Mukohyama00}.) Correspondingly, the action becomes the
following form. 
%============< EQUATION >==============%
%
\begin{equation}
 I = \frac{1}{2}\int dt \left\{ 
	af[\dot{u}_{\perp}^2+k^2(\dot{u}_{\parallel}-u_0)^2]
	+ a^3m^2u_0^2
	-a^{-1}f[k^2+(n-1)K]u_{\perp}^2
	-am^2(u_{\perp}^2+k^2u_{\parallel}^2)\right\}
\end{equation}
%======================================%
The constraint (\ref{eqn:U0}) can be rewritten as
%============< EQUATION >==============%
%
\begin{equation}
 u_0 = \frac{fk^2\dot{u}_{\parallel}}{fk^2+a^2m^2}. 
\end{equation}
%======================================%
By using this expression, we can eliminate $u_0$ in the action to give 
%============< EQUATION >==============%
%
\begin{eqnarray}
 I & = & \int dt (L_{\perp} + L_{\parallel}),\nonumber\\
 L_{\perp,\parallel} & = & 
	\frac{1}{2}\left(\dot{Q}_{\perp,\parallel}^2 
	- \Omega_{\perp,\parallel}^2 Q_{\perp,\parallel}^2\right),
\end{eqnarray}
%======================================%
where $Q_{\perp}$ and $Q_{\parallel}$ are defined by
%============< EQUATION >==============%
%
\begin{eqnarray}
 Q_{\perp} & = & u_{\perp}\sqrt{af},\nonumber\\
 Q_{\parallel} & = & u_{\parallel}
	\sqrt{\frac{afk^2}{1+fm^{-2}a^{-2}k^2}},
\end{eqnarray}
%======================================%
and 
%============< EQUATION >==============%
%
\begin{eqnarray}
 \Omega_{\perp}^2 & = & 
	f^{-1}m^2 + [k^2+(n-1)K]a^{-2} 
	- \Delta_{\perp}^2/2 - \dot{\Delta}_{\perp},
	\nonumber\\
 \Omega_{\parallel}^2 & = & 
	f^{-1}m^2 + k^2a^{-2} 
	- \Delta_{\parallel}^2/2 - \dot{\Delta}_{\parallel}. 
\end{eqnarray}
%======================================%
Here, $\Delta_{\perp}$ and $\Delta_{\parallel}$ are defined by
%============< EQUATION >==============%
%
\begin{eqnarray}
 \Delta_{\perp} & = & \frac{d}{dt}\ln(af),\nonumber\\
 \Delta_{\parallel} & = & \frac{d}{dt}
	\ln\left(\frac{af}{1+fm^{-2}a^{-2}k^2}\right).
\end{eqnarray}
%======================================%
Therefore, hamiltonians for $Q_{\perp}$ and $Q_{\parallel}$ w.r.t. time 
$t$ are 
%============< EQUATION >==============%
%
\begin{equation}
 H_{\perp,\parallel} =  
	\frac{1}{2}\left(P_{\perp,\parallel}^2 
	+ \Omega_{\perp,\parallel}^2 Q_{\perp,\parallel}^2\right),
\end{equation}
%======================================%
where $P_{\perp,\parallel}$ are momenta conjugate to 
$Q_{\perp,\parallel}$, respectively. 

Setting $n=3$, $f=(b/b_0)^d$ and $m^2=(b/b_0)^{-2}m_U^2$, we can apply 
the above result to the Kaluza-Klein mode in
Sec.~\ref{sec:U(1)}. Actually, $\Omega_{\perp,\parallel}^2$ becomes of 
the form. 
%============< EQUATION >==============%
%
\begin{equation}
 \Omega_{\perp,\parallel}^2 =
	m_U^2[1-\epsilon\cos(\omega_bt)] + a^{-2}\tilde{k}^2 
	+ O(\epsilon^2), 
\end{equation}
%======================================%
where $\tilde{k}^2$ is $k^2+2K$ for $\Omega_{\perp}$ and $k^2$ for
$\Omega_{\parallel}$, respectively, and $\epsilon$ is a dimensionless
quantity proportional to the amplitude of oscillation of $b$ around
$b_0$. To derive this expression we have assumed that $Hb_0\ll 1$ and
$\dot{H}b_0^2\ll 1$.

%======================================%
%<<<<<<<<<<<< REFERENCES >>>>>>>>>>>>>>%
%======================================%

\end{document}